\documentstyle[12pt]{article}

\newcommand \be {\begin{equation}}
\newcommand \bea {\begin{eqnarray}}
\newcommand \ee {\end{equation}}
\newcommand \eea {\end{eqnarray}}

\newcommand \dis {\displaystyle}
\newcommand{\for}{\ \ \ \mbox{\rm for}\ \ \ }

\begin{document}

\title{Fluctuations in a Spin Glass model with 1RSB}
\author{M. E. Ferrero, G. Parisi and P. Ranieri\\
{\small Dipartimento di Fisica Universit\`a di Roma La Sapienza}\\ 
{\small and INFN sezione di Roma I}\\ 
{\small Piazzale Aldo Moro, Roma 00185}}
\maketitle

\begin{abstract}
We discuss Gaussian fluctuations in a spin glass model with one replica 
symmetry breaking and we show how non-perturbative fluctuations of the 
break-point parameter can be included in the longitudinal propagator
within linear response theory.  
\end{abstract}

\vfill

The aim of this letter is to discuss the fluctuations and more
generally the corrections to the mean field theory of spin glass models
where first order replica symmetry breaking occurs. We remind 
to the reader that replica symmetry can be broken in two different 
ways \cite{LIBRI}:
\begin{itemize}
          \item  The function $q(x)$ is discontinuous and it takes only 
a finite number of values (in most cases two). Here the function $P(q)$ 
is the sum of a finite number of delta functions. For example in the 
case of only one step (1RSB) we have
\be
 	q_m(x)= q_0 \for x<m, \ \ \ q_m(x)=q_1 \for x>m.
\ee
The corresponding function $P(q)$ is given by
\be
 	P(q)=m \delta(q-q_0)+ (1-m) \delta (q-q_1).
\ee
 	
         \item  The function $q(x)$ is a continuous function and in this 
case also the function $P(q)$ has a continuous part.
\end{itemize}
 
Some models as the Sherrington Kirpatrick and the Edwards Anderson
model belong to the second category, other models, the random energy
model, Ising spins with $p$ interactions and $p>2$, the $q$ state Potts
model with $q>4$, the ROME (random orthogonal matrix ensemble) belong
to the first category.
 
The computation of the fluctuations and the corrections to the
saddle point limit is rather difficult in the second case, where the
form of the propagators is quite involved, and requires many powerful
tools \cite{DKT1}.
 
In the first case (1RSB) the situation was supposed to be much
simpler, the propagator can be explicitly computed taking care of only
the fluctuations of $q_0$ and $q_1$. The problems arise when the
fluctuations in the variable $m$ are considered. 

Fluctuations changing $m$ by a small amount are small in some
sense and they have to be taken into account in the computations, but
in some other sense they are large and the usual formalism (as we shall
see) does not take them into account and must consequently be
modified. Indeed it is
true that when $m \to \tilde{m}$ $q_{m}(x) \to q_{\tilde{m}}(x)$ 
in some sense (for example in the $L^p$ norm with finite $p$),
 but the quantity 

\be
\label{norma}
\sup_{x}( q_{m}(x) - q_{\tilde m}(x))\equiv |q_{m}-q_{\tilde{m}}|_\infty 
\ee 
does not go to zero in this limit.
                                                
One of the  first results suggesting the necessity of taking care of 
fluctuations which correspond to variations of $m$ is the following
\cite{PV}. In the random energy model (REM) of Derrida the free energy 
can be written as
\be
F(\beta)=-\frac{N}{\beta m}\log{2}+ \frac{N \beta m}{2}
\ee
The correct result is obtained as a saddle point in $m$ for large 
$N$. Corrections 
proportional to $1/N$ to the free energy density are clearly connected to 
fluctuations in $m$, while if we consider the formalism of \cite{GM} and 
we represent the REM as a Ising model with a p-spin interaction
these corrections cannot come from fluctuations 
in the $q$ parameters, which vanish in this limit.
 
More recently it has been shown (see \cite{THEO}) that if one does not
take into account the fluctuations of $m$ one obtains the wrong result for
the specific heat while the correct result could be obtained by taking
into account the $m$- fluctuations using a simple (but at this stage
arbitrary) prescription.

The aim of this note is to compute part of the fluctuations (the so
called longitudinal propagator) by using the linear response theory,
i.e. by evaluating the variation of the function $q(x)$ with respect
to an external perturbation. This propagator contains singular terms,
which are not found using the conventional approach.  Correct results
for the specific heat are obtained using this improved propagator.
 
We postpone the computation of the full propagator to a future
investigation.  Here we limit ourselves to the computation of the
longitudinal propagator.

\section{The Model}

We use in our analysis a simple model that we consider
representative of the class of models with a 1RSB saddle point. 
The model is simply obtained by adding an additional cubic term 
to the usual truncated free energy (W in the following), i.e.
\be 
\label{W}
W[Q]=-\lim_{n\rightarrow 0} \frac{1}{n}\left(\frac{\tau}{2} \mbox{Tr} Q^2
+\frac{1}{6} \mbox{Tr} Q^3+ \frac{\alpha}{6}
\sum_{ab}(Q_{ab})^3+\frac{\beta}{12} \sum_{ab}(Q_{ab})^4\right), 
\ee 
where $\tau=T_c-T$ and Tr stands for trace. We recall that in the 
SK model $\alpha=0$ and $\beta=1$ (while, for example, in the 3-state 
Potts model $\alpha=1/2$ and $\beta$ is negative \cite{GKS}). In 
the framework of the
Parisi ansatz the saddle point of $Q$ is looked for in a subspace in
which $Q$ can be expressed in terms of a function $q(x)$ defined in
the interval $[0,1]$. In this subspace the functional $W[Q]$ is given by
\be
\label{W_con}
W [q]=\int_0^1 dx \left[\frac{\tau}{2}q^2(x)
-\frac{1}{6}\left(xq^3(x)+3 q^2(x)\int_x^1 q(y)dy\right)+\frac{\alpha}{6}
q^3(x) +\frac{\beta}{12}q^4(x)\right].  
\ee 
Below $T_c$ stationarity with respect to the order parameter yields the 
1RSB solution
\be
\label{q(x)}
q_m(x) = q_1 \theta(x-m)
\ee
where the parameters $q_1$ and $m$ ($q_0=0$) are obtained 
by the saddle point conditions as a perturbative series in $\beta$, 
\bea 
\label{sol} 
q_1 & \simeq & \frac{\tau}{1-\alpha}+\frac{5}{6}\frac{\beta \tau^2}
{(1-\alpha)^3}+\frac{25}{18}\frac{ \beta^2 \tau^3}{(1-\alpha)^5}\nonumber\\
m & \simeq & \alpha+\frac{\beta \tau}{1-\alpha}+\frac{5}{6}\frac{\beta^2 \tau^2}
{(1-\alpha)^3}.
\eea
The role played by the additional cubic term is to provide a breaking 
of replica symmetry which is located at $m\simeq \alpha$ while it is
well known that in the SK model $m\sim \beta \tau$. To investigate the
stability of this saddle point with respect to $Q$ fluctuations we need
the eigenvalues of the matrix
\be
M_{ab,cd}=\frac{\partial^2 W}{\partial Q_{ab}\partial Q_{cd}}.
\ee
We find that the eigenvalues of this Hessian 
\footnote{For a complete and general analysis of the eigenvectors 
structure in the replica approach, see \cite{DKT2}. In what follows
we use the notation presented in \cite{DKT2}.}, 
which should be positive in order to have a stable saddle point, are
\be
\begin{array}{llll}
\lambda_0 &=\lambda_{1,0}&=-\tau -q_1(m-1) &\rightarrow -\beta q_1^2/6\\     
\lambda_1 &=\lambda_{1,1}&=-\tau -\alpha q_1-\beta q_1^2 -q_1(m-2)
                         & \rightarrow -\beta q_1^2/6+q_1 (1-m)\\          
          &=\lambda_{2,0}&=-\tau -q_1(m/2-1)
                         & \rightarrow -\beta q_1^2/6+q_1 m/2\\     
          &=\lambda_{2,1}&=-\tau -\alpha q_1-\beta q_1^2 -q_1(m/2-2)
                         &  \rightarrow -\beta q_1^2/6+q_1 (1-m/2)\\
          &=\lambda_{0,1,1}&=-\tau-q_1(m-1) & \rightarrow -\beta q_1^2/6\\     
          &=\lambda_{1,2,2}&=-\tau-\alpha q_1-\beta q_1^2+ q_1
                         &   \rightarrow -\beta q_1^2/6\\        
\lambda_{0,1,2}&=\lambda_{0,2,1}&=-\tau-q_1(m/2-1)
                         &      \rightarrow -\beta q_1^2/6+q_1 m/2\\        
         &=\lambda_{0,2,2}&=-\tau+q_1 & \rightarrow -\beta q_1^2/6 +q_1 m.
        
\end{array}
\ee
Using the saddle point values, we find that the minimum eigenvalue belongs 
to four degenerate sub-families ($\lambda_0=\lambda_{1,0}=\lambda_{0,1,1}=
\lambda_{1,2,2}$) and it is proportional to $-\beta \tau^2$. This shows 
that a coefficient $\alpha \neq 0$ allows the prescription of keeping 
a negative coefficient $\beta$ in order to have a stable 1RSB ansatz 
(see also \cite{FV}), without a negative value for $m$ and without a 
negative eigenvalue. 

\section{Fluctuations}

Let us now consider this model in the Gaussian approximation.  
Our aim is to derive, within linear response theory, a longitudinal 
propagator which takes into account $m$ fluctuations. 
In order to have a consistent check of our computation, at the end of 
this section we shall compare the specific heat obtained through
this improved propagator with the usual expression obtained through 
the saddle point solution. 

In the replica approach, the longitudinal propagator can be computed
in the discrete formulation of replica symmetry breaking, i. e.  
by using the parameters $q_0$, $q_1$ and $m$ and the global variations 
$\delta q_0$, $\delta q_1$ and $\delta m$ (as done in \cite{THEO}), or 
by considering the function $q(x)$ (see Eq. (\ref{W_con})) and the local 
variations $\delta q(x)$ (eventually followed by integration).
Clearly, if we work in the local formulation, that is the first
step toward the analysis in the full space,  
we need a method to deal with the $m$ fluctuations. 
As previously mentioned, these fluctuations induce a non-perturbative 
(not small in the sense of Eq. (\ref{norma})) variations on the function 
$q(x)$ and it is unclear how they can be taken into account in a perturbative
computation. 

In order to take into account these fluctuations let us introduce 
in this model an external field conjugate with the order parameter:
\be
W[q] \rightarrow W[q] +\int_0^1 q(x) \epsilon (x).
\ee
The perturbation induced by this field on the saddle 
point solution can be parametrized as follows:
\be
\label{q_e}
q^{\epsilon}_m(x)=q_1 \theta(x-m-\delta m)+\delta q(x).
\ee
Within linear response theory, we define the longitudinal propagator
by considering the response with respect to $\epsilon$, i. e. 
\be
\label{prop}
G(x,y)=\frac{\delta q^{\epsilon}_m(x)}{\delta \epsilon (y)}=     
-q_1 \delta (x-m) \frac{\delta m}{\delta \epsilon (y)}
+\frac{\delta q(x)}{\delta \epsilon (y)}.
\ee

The equations for the two components of the propagator  
follow from the equations of motion with the source $\epsilon \neq 0$   
and from their expansion to first order in $\epsilon$, i. e.
\bea
\label{eq1}
\left.\frac{\delta W[q]}{\delta q(x)}\right|_{q=q_{\epsilon}}&=&
\epsilon(x)\\
\label{eq2}
\frac{\partial W[q_{\epsilon}]}{\partial m}&=&-q_1\epsilon(m).
\eea
Using this procedure we manage to 
consider in a perturbative approach a non-perturbative contribution.
On the one hand the variations $\delta q(x)$ 
and $\delta m$ defined in Eq. (\ref{q_e},\ref{prop}) play a different role. 
The introduction of $\delta(x-m)$ as a multiplicative factor of the component 
$\frac{\delta m}{\delta \epsilon (y)}$ is crucial in (\ref{prop}) because 
this delta-function separates without ambiguities the two contributions.
On the other hand the two equations (\ref{eq1}, \ref{eq2}) are 
qualitatively
different: the first is a functional derivative of the free energy functional 
$W[q]$ while the latter is a derivative of a function $W(q_{\epsilon}(x))=
\tilde{W}_{q_{1},m}(\delta q(x), \delta m)$ with respect to $\delta m$. 

There, while in the Eq. (\ref{eq2}) the distribution functions are 
integrated and no ambiguity exists, in the Eq. (\ref{eq1}) we have to deal 
with products of distribution functions (i.e. $\theta^2 (x-m)$, 
$\theta(x-m)\delta(x-m)$ and $\theta^2(x-m)\delta(x-m)$). These products are, 
at this stage, ill defined and a regularization scheme is necessary. In what
follows we choose a regularization such that 
\bea
\label{def}
&\theta^k (x-m)=\theta (x-m),& \\
\label{def1}
&\theta^{k-1}(x-m) \delta(x-m)=\frac{1}{k}\delta(x-m)&,
\eea  
where the function $\delta (x-m)$ that occurs in  
Eq. (\ref{prop}) is defined as the derivate of the function $\theta(x-m)$.
Therefore relation (\ref{def1}) is the derivative of 
relation (\ref{def}), that is the only arbitrary choice we make.
One can also see that Eq. (\ref{def1}) involves the following
prescription to evaluate  the integral of the function $q^k(x)$ on a peaked 
measure:                     
\be
\label{def2}
\int_0^1 q(x)^{k} q_{1}\delta(x-m)dx = \int_{0}^{q_1}q^{k}dq=
\frac{q_1^{k+1}}{k+1}.
\ee
                                          
By expanding equations (\ref{eq1}) and (\ref{eq2}) to first order 
in $\epsilon$ and by using (\ref{def}) and (\ref{def1}) we obtain following 
equations for the propagator components: 
\bea                                               
&\dis{ -q_{1}\theta(x-m)\int_{m}^{1}dy 
\frac{\delta q(y)}{\delta \epsilon(z)}
+\frac{1}{6} \beta q_{1}^2 
\frac{\delta q(x)}{\delta \epsilon(z)}
+\frac{1}{2} q_1^2 \theta (x-m) \frac{\delta m}{\epsilon (z)} = 
\delta (x-z) }& \nonumber\\
&\dis{ \frac{1}{2}q_1^2 \int_m^1 dy 
\frac{\delta q(y)}{\delta \epsilon (z)} 
-\frac{1}{3} q_1^3 \frac{\delta m}{\delta \epsilon (z)} = 
-q_1 \delta (m-z). }&
\eea
The corresponding result for the longitudinal propagator (\ref{prop}) is
\bea
\label{propagatore}
G(x,y)&=&G_0\; \delta (x-y)
+G_1\; \theta(x-m)\theta(y-m) +
\nonumber \\
  &-&\left(G^{N}_0 \delta(x-m)+ G^{N}_1 \theta(x-m)\right) 
\left(G^{N}_0 \delta(y-m)+ G^{N}_1\theta(y-m)\right) 
\eea
where
\bea
&&
G_0=\frac{1}{q_{1}^2\,\beta/6},\ \  
G_1=\frac{1}{q_{1}^2\,\beta/6}\frac{q_{1}}{(q_1^2\,\beta/6 -(1-m)\,q_1)},\ \
G^N_0=\sqrt{\frac{3(q_1^2\,\beta/6-(1-m)\,q_1)}{q_1 \,
(q_1^2\,\beta/6-(1-m)\,q_1/4)}},
\nonumber\\
&&
G^N_1=\sqrt{\frac{3/4}{(q_1^2\,\beta/6 -(1-m)\,q_1/4) 
(q_1^2\,\beta/6 -(1-m)\,q_1)}}.
\eea
Two new terms, overlooked by the usual computation, appear in this 
longitudinal propagator, the term $G_0^N$ and $G_1^N$. These terms, singular 
at $x\simeq m$, are the effect of the $m$ fluctuations. Let us also note that 
the asymmetry between the regions $x>m$ and $x<m$ in this result is 
due to the assumption $q_{0}=0$ on the saddle point.

To conclude, let us verify the previous result and let us investigate
its consequence on physical quantities, as the specific heat. 
It is well known that this quantity can be computed through the free energy 
evaluated at the saddle point or by computing the energy-energy 
fluctuations \cite{giorgiolibro}. 
The computation of a specific heat in the mean field approximation
through the 1RSB saddle point gives
\be
\label{calspec}
C(\tau)=-\frac{d^2}{d \tau^2} W[q]_{SP}= -\frac{\tau}{1-\alpha}- 
\frac{\beta \tau^2}{(1-\alpha)^3}-\frac{35}{18}\frac{\beta^2\tau^3}
{(1-\alpha)^5},
\ee
where the dependence of the $m$ parameter on the temperature 
implies a contribution to the specific heat also from the variation of $m$. 

On the other hand, by considering the Gaussian fluctuations  
at zero-loop order and by using our prescriptions to
deal with the distribution functions, we find also that
\be
\label{calspecbis}
C(\tau)= \frac{1}{4} \langle \int dx\; dy\; q^2(x)\;q^2(y)\rangle _{conn}=
\int_0^1 dx \int_0^1 dy \; q_{SP}(x) q_{SP}(y) G(x,y).
\ee

This shows that the new singular terms, which with our prescription 
(\ref{def2}) produce an effect in the computation of the physical 
quantity (\ref{calspecbis}), are necessary to recover the correct result.
Because of the nature of the $x$ variable in the replica approach, 
the prescriptions for the singular measures are necessary to recover the
correct result, while in the discrete formalism, where one has to deal 
with the parameters $q_1$ and $m$ only, the regularization is not 
necessary and one naturally recovers the correct result.                     

In the case of a continuous breaking of the replica symmetry the
longitudinal propagator computed using the linear response theory do
coincide with the one obtained by the conventional approach \cite{FP}.  
Our result therefore suggests the following scenario
\begin{itemize} 

\item If we break the replica symmetry in a continuous way by adding
an appropriate external field, the longitudinal propagator is
correctly given by the conventional techniques.

\item If, by removing the external field, the function $q(x)$ becomes
discontinuous, the longitudinal propagator computed via linear
response theory goes to the correct one and therefore also the
conventionally computed propagator will tend to the same value, which
is different from the value obtained by applying directly the
conventional techniques.
 
\item We may only conjecture that a correct computation of all the
 	components of the propagator (not only the longitudinal one)
 	may be achieved by using the conventional approach after having
 	introduced an external field which breaks the replica symmetry
 	in a continuous way and then by sending the external
 	field to zero.  
\end{itemize}

\vspace{0.5cm}

It is a pleasure to thank Theo Nieuwenhuizen for communicating his 
results before publication and for useful discussion. We also thank 
David Dean for a careful reading of the manuscript.

\end{document}